\documentclass[12pt,epsf]{article}
\usepackage{epsf}
\usepackage{graphicx}
\begin{document}

\def\a{\alpha}
\def\b{\beta}
\def\c{\varepsilon}
\def\d{\delta}
\def\e{\epsilon}
\def\f{\phi}
\def\g{\gamma}
\def\h{\theta}
\def\k{\kappa}
\def\l{\lambda}
\def\m{\mu}
\def\n{\nu}
\def\p{\psi}
\def\q{\partial}
\def\r{\rho}
\def\s{\sigma}
\def\t{\tau}
\def\u{\upsilon}
\def\v{\varphi}
\def\w{\omega}
\def\x{\xi}
\def\y{\eta}
\def\z{\zeta}
\def\D{\Delta}
\def\G{\Gamma}
\def\H{\Theta}
\def\L{\Lambda}
\def\F{\Phi}
\def\P{\Psi}
\def\S{\Sigma}

\def\o{\over}
\def\beq{\begin{eqnarray}}
\def\eeq{\end{eqnarray}}
\newcommand{\gsim}{ \mathop{}_{\textstyle \sim}^{\textstyle >} }
\newcommand{\lsim}{ \mathop{}_{\textstyle \sim}^{\textstyle <} }
\newcommand{\vev}[1]{ \left\langle {#1} \right\rangle }
\newcommand{\bra}[1]{ \langle {#1} | }
\newcommand{\ket}[1]{ | {#1} \rangle }
\newcommand{\EV}{ {\rm eV} }
\newcommand{\KEV}{ {\rm keV} }
\newcommand{\MEV}{ {\rm MeV} }
\newcommand{\GEV}{ {\rm GeV} }
\newcommand{\TEV}{ {\rm TeV} }
\def\diag{\mathop{\rm diag}\nolimits}
\def\Spin{\mathop{\rm Spin}}
\def\SO{\mathop{\rm SO}}
\def\O{\mathop{\rm O}}
\def\SU{\mathop{\rm SU}}
\def\U{\mathop{\rm U}}
\def\Sp{\mathop{\rm Sp}}
\def\SL{\mathop{\rm SL}}
\def\tr{\mathop{\rm tr}}

\def\IJMP{Int.~J.~Mod.~Phys. }
\def\MPL{Mod.~Phys.~Lett. }
\def\NP{Nucl.~Phys. }
\def\PL{Phys.~Lett. }
\def\PR{Phys.~Rev. }
\def\PRL{Phys.~Rev.~Lett. }
\def\PTP{Prog.~Theor.~Phys. }
\def\ZP{Z.~Phys. }

\def\stau{\widetilde{\tau}}


\newcommand{\gtrsim}{ \mathop{}_{\textstyle \sim}^{\textstyle >} }
\newcommand{\lesssim}{ \mathop{}_{\textstyle \sim}^{\textstyle <} }
\newcommand{\rem}[1]{{\bf #1}}

\baselineskip 0.7cm

\begin{titlepage}

\begin{flushright}
SLAC-PUB-12168\\
TU-778\\
October, 2006
\end{flushright}

\vskip 1cm
\begin{center}
{\Large \bf
Possible Signals of Wino LSP at the Large Hadron Collider
}

\vskip 1.2cm

M. Ibe${}^{1}$, Takeo Moroi${}^{2}$ and T. T. Yanagida${}^{3,4}$

\vskip 0.4cm

${}^{1}${\it Stanford Linear Accelerator Center, Stanford University, \\Stanford, CA94309}

\vskip 0.3cm

${}^2${\it Department of Physics, Tohoku University,\\
     Sendai, Japan}

\vskip 0.3cm

${}^3${\it Department of Physics, University of Tokyo,\\
     Tokyo 113-0033, Japan}

\vskip 0.3cm

${}^4${\it Research Center for the Early Universe, University of Tokyo,\\
     Tokyo 113-0033, Japan}

\vskip 1.5cm

\abstract{

We consider a class of anomaly-mediated supersymmetry breaking models
where gauginos acquire masses mostly from anomaly mediation while
masses of other superparticles are from K\"ahler interactions, which
are as large as gravitino mass $\sim{\cal O}(10-100)$ TeV.  In this
class of models, the neutral Wino becomes the lightest superparticle
in a wide parameter region.  The mass splitting between charged and
neutral Winos are very small and experimental discovery of such Winos
is highly non-trivial.  We discuss how we should look for Wino-induced
signals at Large Hadron Collider.

 }
\end{center}

\end{titlepage}

\setcounter{page}{2}

\section{Introduction}

In the vacua of broken supersymmetry (SUSY) not only the gravitino but
also squarks and sleptons acquire tree-level SUSY-breaking masses
through supergravity (SUGRA) effects~\cite{SUSY}. On the other hand,
if there is no singlet field in the hidden sector, quantum-loop
corrections generate gaugino masses, which is often called as anomaly
mediation~\cite{Randall:1998uk,Giudice:1998xp}. This is the most
economical mechanism for giving SUSY-breaking masses to all SUSY
particles in the SUSY standard model (SSM) since it does not require
any extra field other than ones responsible for the SUSY breaking. In
this model the gauginos are much lighter than the squarks and sleptons
due to the loop effects. That is, there is a little disparity in the
spectrum of SUSY particles. The masses of squarks and sleptons (and
also that of the gravitino) are ${\cal O}(10-100)$ TeV for the gaugino
masses less than ${\cal O}(1)$ TeV. Because of this disparity the
anomaly mediation model is free from all problems in the SSM such as
flavor-changing neutral current and CP problems~\cite{FCNC}.
Furthermore, the mass of the lightest Higgs boson is naturally above
the experimental lower bound $m_H> 114.4$ GeV~\cite{Yao:2006px}.  From
the cosmological point of view, the gravitino-overproduction
problem~\cite{gravitino} in the early universe is much less serious in
the above model due to the relatively large gravitino mass.  It is
also notable that a consistent scenario of thermal leptogenesis
\cite{Fukugita:1986hr} is possible in the present scheme
\cite{Leptogen+AMSB}, and that the lightest superparticle (LSP), which
is the neutral Wino, can be a good candidate for the dark matter in
the universe \cite{Moroi:1999zb,ISY}.

The purpose of this paper is to discuss possible tests of the above
anomaly mediation model at the Large Hadron Collider (LHC)
experiments. The crucial point is that the charged and neutral Winos
are the most-motivated LSPs whose masses almost degenerate with each
other, and the neutral Wino $\tilde{W}^0$ is slightly lighter
than the charged one $\tilde{W}^\pm$ \cite{Feng:1999fu}.
In a cosmological scenario \cite{ISY}, a sizable amount
of Winos are produced by the decay of gravitino originating from
inflaton decay \cite{Kawasaki:2006gs}, and provided that such Wino can be 
the dark matter in the universe, 
the mass of the Wino is predicted as $M_2\simeq100 \,{\rm GeV}-2$\,TeV.
Therefore, it is highly possible that the Wino is within the reach of the LHC.%
\footnote
{The thermally produced Winos cannot explain the observed dark matter
density unless it is very heavy, $M_{2}\simeq 2$\,TeV, which requires
gravitino with mass $m_{3/2}= {\cal O}(1)$\,PeV~\cite{Wells:2004di}. }

Unfortunately, however, it is also possible that 
superparticles other than Winos can be hardly produced at the LHC,
depending on the model parameters.
Indeed, as we will see, the gluino-Wino mass ratio $M_{3}/M_{2}$
can be larger than the prediction of the pure anomaly mediation model,
$M_{3}/M_{2}\simeq 8$. 
Thus, it is difficult to produce the gluinos 
at the LHC except for very optimistic cases as well as sfermions. 
Therefore it is now highly important to discuss signals of the Wino production at the LHC.

\section{Properties of Winos}

We first discuss the mass spectrum of superparticles.  We assume that
the masses of squarks, sleptons and Higgs bosons are dominantly from
SUGRA effects.  With a generic K\"ahler potential, all the scalar
bosons except the lightest Higgs boson acquire SUSY breaking masses of
the order of the gravitino mass ($m_{3/2}$).  For the gaugino masses,
on the contrary, tree-level SUGRA contributions are extremely
suppressed if there is no singlet elementary field in the hidden
sector.  In this paper, we adopt the above setup and consider the
situation that the scalar masses (as well as the gravitino mass) are
of the order of ${\cal O}(10-100)$ TeV while gauginos are much
lighter.  In addition, we also assume that Higgsinos and heavy Higgs
bosons have masses of ${\cal O}(10-100)$ TeV.  Hereafter, we call such
a scenario as ``no singlet scenario''.

If there is no singlet field in the hidden sector, effects of anomaly
mediation becomes very important in generating gaugino masses
\cite{Randall:1998uk,Giudice:1998xp}.  At $Q\sim m_{3/2}$ (with $Q$
being a renormalization scale), anomaly-mediation contributions to the
gaugino masses are given by
\begin{eqnarray}
    M_{a}^{(AMSB)}
    = -\frac{b_{a}g_{a}^{2}}{16\pi^{2}} m_{3/2},
    \label{eq:GauginoMass}
\end{eqnarray}
where $a$ runs in three standard-model gauge groups ($a=1,2,3$),
$g_{a}$ denote gauge coupling constants, and $b_{a}$ coefficients of
the renormalization-group equations for $g_{a}$, i.e., $b_{a}= (-33/5,
-1, 3)$.

The important point is that the natural sizes of so-called $\mu$- and
$B$-parameters for Higgs multiplets (as well as their SUSY breaking
masses) are as large as the gravitino mass in the ``no singlet
scenario''.  Indeed, we expect the following K\"ahler potential:
\begin{eqnarray}
    K \ni c H_{u}H_{d} + 
    \frac{c'}{M_{G}^{2}} X^{\dagger} X H_{u}H_{d} + h.c.,
\end{eqnarray}
where $X$ denotes a chiral superfield in the hidden sector, which may
be elementary or composite, $M_G$ is the reduced Planck scale, and $c$
and $c'$ are coefficients of ${\cal O}(1)$.  Then, the $\mu$- and the
$B$-parameters are given by \cite{Inoue:1991rk}\footnote
{We assume that the vacuum expectation value of $X$ is much smaller
than $M_G$.}
\begin{eqnarray}
    \label{eq:Muterm}
    \mu &=& c m_{3/2},\\
    \label{eq:Bterm}
    B \mu &=& c m_{3/2}^{2} + c'\frac{|F_X|^2}{M_{G}^{2}},
\end{eqnarray}
where $F_X$ is the vacuum expectation value of the $F$-component of
$X$.  Thus, $\mu$- and $B$-parameters are both of ${\cal O}(m_{3/2})$
and they are independent.  For the successful electro-weak symmetry
breaking, one linear combination of Higgs bosons should become light:
with the so-called $\beta$-parameter, the (standard-model-like) light
doublet is denoted as $\sin\b H_{u} - \cos\b H_{d}^{*}$, where $H_u$
and $H_d$ are up- and down-type Higgs bosons, respectively.  

Then, threshold corrections to the gaugino masses from the
Higgs-Higgsino loop are given by
\cite{Giudice:1998xp,Gherghetta:1999sw}
\begin{eqnarray}
    \Delta M_{1}^{(Higgs)} &=& 
    \frac{3}{5} \frac{g_{1}^{2}}{16 \pi^{2}} L,
    \label{DelM1} \\
    \Delta M_{2}^{(Higgs)} &=& 
    \frac{g_{2}^{2}}{16 \pi^{2}} L,
    \\
    \Delta M_{3}^{(Higgs)} &=&  0,
    \label{DelM3}
\end{eqnarray}
where
\begin{eqnarray}
    L \equiv \mu \sin2\beta 
    \frac{m_{A}^{2}}{|\mu|^{2}-m_{A}^{2}} \ln \frac{|\mu|^{2}}{m_{A}^{2}}.
    \label{L-parameter}
\end{eqnarray}
Here, $m_A$ is the mass of heavy Higgs bosons. 
When $\mu={\cal O}(m_{3/2})$, $\Delta M_{a}^{(Higgs)}$ (for $a=1,2$)
and $M_{a}^{(AMSB)}$ are of the same order, and hence we have sizable
deviations from the relations in Eq.\ (\ref{eq:GauginoMass}) of the
pure anomaly mediation. 

\begin{figure}[t]
    \centerline{\epsfxsize=0.6\textwidth\epsfbox{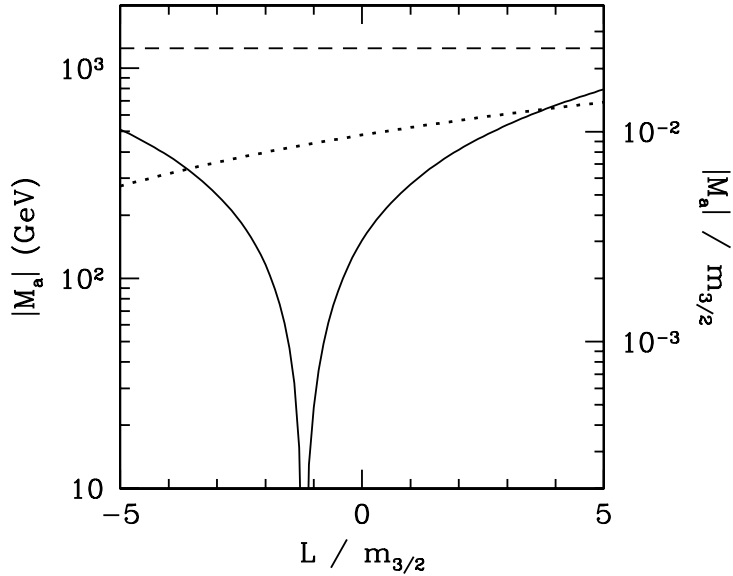}}
    \caption{Gaugino masses as functions of $L/m_{3/2}$ for a real value of $L$.
    The Wino  mass is given in the solid line, while the dotted and dashed lines
    are for Bino and gluino masses, respectively.  Here we take
    $m_{3/2}=50$ TeV. 
    The cancellation of the Wino mass for $L/m_{3/2}\simeq -1$ is a result of our specific choice of 
    the phase, Arg$(L)=\pi$.
    For a generic phase of $L$, the cancellation becomes milder, though the Wino mass can be 
    much lighter than $M_{2}^{(AMSB)}$ for $|L/m_{3/2}|\simeq 1$ and Arg$(L)\sim\pi$.}
    \label{fig:Ma}
\end{figure}

The $\mu$-parameter (and $L$) is a complex variable and the Wino and
Bino masses depend on the relative phase between $\mu$ and $m_{3/2}$.
(Here, we use the bases where the gravitino mass is real and positive.)
Importantly, with a distractive interference between $M_2^{(AMSB)}$
and $\Delta M_2^{(Higgs)}$, the Wino mass can be even smaller than the
purely anomaly mediated one.  As an example, in Fig.\ \ref{fig:Ma}, we
plot the gaugino masses for $m_{3/2}=50$ TeV as functions of
$L/m_{3/2}$, assuming that $L$ is real for simplicity.  In
this case, we see that the threshold corrections drastically change
the gaugino mass and even cancel the anomaly mediated Wino mass.  In
our analysis, gaugino masses are given by $M_a=M_a^{(AMSB)}+\Delta
M_a^{(Higgs)}$ at $Q=m_{3/2}$, and we take into account the effects of
renormalization group evolutions below this scale.  We have checked
that the ratios of two gaugino masses are insensitive to the gravitino
mass for a given $L/m_{3/2}$.

A wide variety of gaugino mass in the ``no singlet scenario'' has strong
implications for collider physics. 
 In the pure anomaly mediation, all the gaugino masses are
determined only by the gravitino mass and corresponding gauge coupling
constants as shown in Eq.\ (\ref{eq:GauginoMass}), and hence a
negative search for gluino-induced signals results in lower bounds on
other gaugino masses, in particular, that of Wino mass.  In the ``no
singlet scenario'', on the contrary, gluino and Wino masses are
independent.  Thus, even if gluino-induced signals will not be found
at the LHC, we will still have a strong motivation to look for
Wino-induced signals for the parameter region where Wino is relatively
light.

In the pure anomaly mediation model, the lightest gaugino is always
the Wino.  However, in the ``no singlet scenario'', the Wino LSP is
not a generic consequence any more since the Bino may become lighter
than Wino.  However, Fig.\ \ref{fig:Ma} shows that Wino becomes the
LSP if $|L/m_{3/2}|$ is less than a few, which is the case when
$|\mu|$ and $m_A$ are both fairly close to the gravitino mass.  In
addition, the Wino LSP scenario can never cease from being the most
motivated scenario from the point of view of cosmology.  This is
because the Bino LSP results in the overclosure of the
universe~\cite{Giudice:1998xp}.
Therefore, in this paper, we concentrate on the Wino LSP scenario,
where the lightest neutralino and the lightest chargino are both
(almost) purely Winos.

Now, we consider detailed properties of the Wino.  As discussed in
Ref.~\cite{Feng:1999fu}, the dominant mass splitting between the
charged and neutral Winos comes from one-loop gauge boson
contributions to the gaugino masses when $\mu$ is large.  The
splitting is given by
\begin{eqnarray}
{\mit \D} M =  m_{{\tilde W}^{\pm}}-m_{{\tilde W}^{0}}
=\frac{g_{2}^{2}}{16\pi^{2}} M_{2} 
\left[ f(r_{W})-\cos^{2}\h_{W} f(r_{Z})-\sin^{2}\h_{W} f(0) \right],
\label{deltaM}
\end{eqnarray}
where $f(r)= \int^{1}_{0} dx(2 + 2 x^{2}) \ln[x^{2} + (1-x)r^{2}]$ and
$r_{i}=m_{i}/M_{2}$.  Requiring that the charged Wino mass be larger
than the current experimental bound $M_{2}\geq 88$ GeV
\cite{Heister:2002mn}, the mass difference is in the range $155\ {\rm
MeV}\lesssim {\mit \D} M \lesssim 170\ {\rm MeV}$.  Then, the dominant
decay mode of charged Wino is given by ${\tilde W}^{\pm}\to {\tilde
W}^{0}\pi^{\pm}$; the decay rate of this process is
\begin{eqnarray}
    \Gamma({\tilde W}^{\pm}\to {\tilde W}^{0} \pi^{\pm}) = 
    \frac{2 G_{F}^{2}}{\pi} \cos^{2}\h_{c} f_{\pi}^{2}
    {\mit \D} M^{3} 
    \left(1- \frac{m_{\pi}^{2}}{{\mit\D}M^{2}}\right)^{1/2},
    \label{WinoWidth}
\end{eqnarray}
where $f_{\pi} \simeq 130$\,MeV, and $\h_{c}$ is the Cabbibo angle.
Notice that the three-body decay processes ${\tilde W}^{\pm}\to
{\tilde W}^{0} l^{\pm}\n_{l}$ are negligible in our study; branching
ratios of this type of processes are a few \% or smaller for
$M_{2}\gsim 88$\,GeV.\footnote
{See the related calculation of the branching ratios for a
nearly degenerate heavy lepton pair in Ref.~\cite{Thomas:1998wy}.}
We also estimate the lifetime of charged Wino, which is given by
${\cal O}(10^{-10})$ sec.  Then, charged Winos produced at the LHC
travel typically ${\cal O}(1-10)$ cm before they decay.  As we
discuss in the following, this fact has very important consequences in
the study of our scenario at the LHC.

\section{Wino at the LHC}

Now we are at the position to discuss how we can test the ``no singlet
scenario'' at the LHC.  In studying superparticles at the LHC, it is
often assumed that the productions of superparticles are mostly via
productions of colored superparticles.  Even though the primary
superparticles are scalar quarks and/or gluino, they decay into
various lighter superparticles which may be scalar leptons, charginos,
and/or neutralinos.  Of course, those processes are important when
scalar quarks and/or gluino are not too heavy.

Since we are interested in the case where all the sfermions (as well
as heavy Higgs multiplets) have masses of order $100\ {\rm TeV}$, they
are irrelevant for the LHC.  Hereafter, we will not consider the
production of those particles.

On the contrary, masses of gauginos (in particular, that of gluino)
are of order 100 GeV -- 1 TeV.  As discussed in the previous section,
the gluino-to-Wino mass ratio is a free parameter in the ``no singlet
scenario'', since the prediction of pure anomaly mediation, which
gives $M_3/M_2\simeq 8$, can be significantly altered.  For a given
Wino mass $M_2$, the discovery of SUSY at the LHC depends on the ratio
$M_3/M_2$.

When $M_3/M_2$ is relatively small, the gluino is light enough so that
a significant amount of gluino is produced at the LHC.  In this case,
the dominant production processes of superparticles are gluino pair
productions.  The primary gluinos decay into lighter superparticles
and, at the end of decay chain, the neutral Winos are produced.
Consequently, we observe events with energetic jets and large missing
$E_T$.  In this case, signals from the production of superparticles
are basically the same as those in well-studied SUSY models.

Hereafter, we concentrate on a rather pessimistic case where the gluino
mass is so large that a gluino production rate is suppressed at the
LHC.  In this case, the only possibility of detecting superparticles
is direct productions of charged and neutral Winos.  (Notice that the
Bino production cross section is very small since the scalar quarks
are extremely heavy.)

In our case, the neutral Wino is the LSP and hence is stable.  Thus,
even if it is produced at the LHC, it does not leave any consequence
on the detector.  The charged Wino is heavier than the neutral one and
hence is unstable. The mass difference $\mit\Delta M=m_{{\tilde
W}^\pm}-m_{{\tilde W}^0}$ is expected to be 160 -- 170 MeV, and the
charged Wino mainly decays as ${\tilde W}^\pm\rightarrow {\tilde
W}^0\pi^\pm$ with lifetime of ${\cal O}(10^{-10})$ sec.  Thus, most of
charged Winos decay before it reaches a muon detector.  In addition,
the emitted pion has very tiny energy; its boost factor is typically
$\gamma_\pi\sim {\cal O}(1)$.  Thus, it is challenging to identify
such a low-energy pion.  These facts make the discover of Wino
production events very difficult.

Winos can be pair-produced at the LHC via Drell-Yang processes
$q\bar{q}\rightarrow\tilde{W}^+\tilde{W}^-$ and
$q\bar{q}'\rightarrow\tilde{W}^\pm\tilde{W}^0$.  However, these events
cannot be recorded since there is no trigger relevant for them.
Indeed, in order for the event to be recorded, the final state should
contain high energy jet(s) and/or high energy electro-magnetic
activity.  In the Drell-Yang process, we do not expect such activities
since the charged Winos decay before reaching the muon detector.

However, events may be recorded if Winos are produced with high $E_T$
jet(s).  In particular, ATLAS \cite{ATLAS} and CMS \cite{CMS} both
plan to trigger on missing $E_T$ events with energetic jets; if the
transverse energy of the jet and missing $E_T$ are both larger than 50
-- 100 GeV, events will be recorded.  (Hereafter, we call such a
trigger as J+XE trigger.)

With $pp$ collision at the LHC, a Wino pair can be produced in
association with a high $E_T$ jet via the following parton-level
processes:
\begin{eqnarray}
    &&
    q\bar{q} \rightarrow \tilde{W}^+ \tilde{W}^- g,~~~
    gq \rightarrow \tilde{W}^+ \tilde{W}^- q,~~~
    g\bar{q} \rightarrow \tilde{W}^+ \tilde{W}^- \bar{q},
    \label{w+w-j}
    \\ &&
    q\bar{q}' \rightarrow \tilde{W}^\pm \tilde{W}^0 g,~~~
    gq \rightarrow \tilde{W}^\pm \tilde{W}^0 q',~~~
    g\bar{q} \rightarrow \tilde{W}^\pm \tilde{W}^0 \bar{q}'.
    \label{wpmw0j}
\end{eqnarray}
Gluon and (anti-)quark in the final state are hadronized and become
energetic jets.  Thus, at the trigger level, this class of events are
characterized by a high $E_T$ mono-jet, missing $E_T$, and no other
energetic activity, since the charged Wino decays before reaching the
muon detector.

\begin{figure}[t]
    \centerline{\epsfxsize=0.5\textwidth\epsfbox{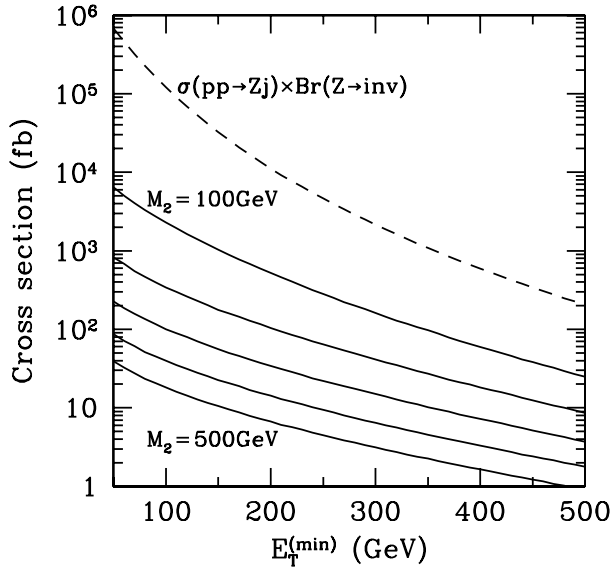}}
    \caption{Total cross section of a
    Wino pair $+$ mono-jet production event $\sigma(pp\rightarrow
    \tilde{W}^+\tilde{W}^-j)+
    \sigma(pp\rightarrow\tilde{W}^\pm\tilde{W}^0j)$ at $\sqrt{s}=14$
    TeV as a function of $E_T^{(min)}$ (solid).  $M_2$ is taken to be
    100, 200, 300, 400, and 500 GeV from above.  Cross section of the
    process $pp\rightarrow Zj$ multiplied by the invisible branching
    ratio of $Z$ is also plotted (dashed).}
    \label{fig:cs_et}
\end{figure}

In order to estimate the total cross section of the Wino pair $+$
mono-jet production process, we calculate the cross section of the
processes (\ref{w+w-j}) and (\ref{wpmw0j}).  We require that the
transverse energy of mono-jet be larger than $E_T^{(min)}$, and
calculate the cross section as a function of $E_T^{(min)}$.  In our
calculation, $E_T$ of jet is approximated to be the same as that of
the final-state gluon or (anti-)quark.  We have used the CTEQ6L parton
distribution function \cite{Pumplin:2002vw}.  The result is shown in
Fig.\ \ref{fig:cs_et}.%
\footnote{
We have cross checked our results by using the codes, {\it FeynArts}~\cite{FA} and 
{\it FormCalc}~\cite{FC}.
}
 With the luminosity of ${\cal O}(10-100)\ {\rm
fb}^{-1}$, sizable numbers of Wino production events are expected even
if we require that the associated jet has transverse energy of ${\cal
O}(100)$ GeV.

Once the events are recorded, the processes
$pp\rightarrow\tilde{W}^+\tilde{W}^-j$ and
$pp\rightarrow\tilde{W}^\pm\tilde{W}^0j$ (with $j$ being jet) may
provide distinguishable signals of the Wino LSP scenario at off-line
analysis.  One possible signal is the track of charged Wino, as also
suggested for the study of Tevatron \cite{Feng:1999fu}.  If the
charged Wino travels ${\cal O}(10)$ cm or longer before it decays, it
hits some of the detectors and the track from charged Wino may be
reconstructed.  
The track from charged Wino may be also distinguished from
tracks from other standard-model charged particles by using the
time-of-flight information and/or by measuring ionization energy loss
rate ($dE/dx$).

In addition, it is important to note that, with the lifetime of ${\cal
O}(10^{-10})$ sec, most of the charged Winos are likely to decay
inside the detector.  Then, we may observe tracks which disappear
somewhere in the detector.  This will be regarded as a spectacular
signal of new physics beyond the standard model.  In particular, in
the ATLAS detector, the transition radiation tracker (TRT) will be
implemented, which is located at 56 -- 107 cm from the beam axis~\cite{ATLASin}.
 The
TRT continuously follows charged tracks.  Thus, if the charged Wino
decays inside the TRT, the charged Wino track will be identified by
off-line analysis.  In addition, with the TRT, charged pions emitted
by the decays of $\tilde{W}^\pm$ may be also seen.  Such events have
very rare backgrounds and hence they can be used for the discovery of
the Wino production.

We calculate the cross section of the processes
$pp\rightarrow\tilde{W}^+\tilde{W}^-j$ and
$pp\rightarrow\tilde{W}^\pm\tilde{W}^0j$, with the requirement that at
least one charged Wino travels transverse length $L_T$ longer than
$L_T^{(min)}$ before it decays.  At the trigger level, no hadronic nor
electro-magnetic activities will be identified except for the
mono-jet, so the missing $E_T$ is equal to the transverse energy of
the mono-jet.  In order to use the J+XE trigger, we impose a cut such
that $E_T\geq 100$ GeV for the final-state jet.  $L_T$ is sensitive to
the decay width of charged Wino.  Since we consider the case where
gauginos are the only light superparticles, we use Eq.\ (\ref{deltaM})
for $\Delta M$.  The decay width of charged Wino is given in Eq.\ 
(\ref{WinoWidth}).

\begin{figure}[t]
    \centerline{\epsfxsize=0.5\textwidth\epsfbox{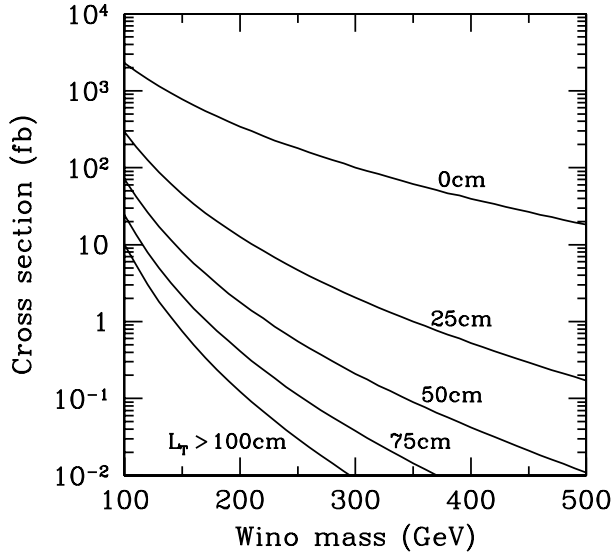}}
    \caption{Cross section of the process
    $pp\rightarrow\tilde{W}\tilde{W}j$ at $\sqrt{s}=14$ TeV as a
    function of $M_2$.  $E_T$ of the jet is required to be larger than
    100 GeV, and the minimum value of $L_T$ is 0, 25, 50, 75, and 100
    cm.}
    \label{fig:cs_ltdep}
\end{figure}

Cross sections for several values of $L_T^{(min)}$ are shown in Fig.\ 
\ref{fig:cs_ltdep}.  Assuming ${\cal L}\sim 100\ {\rm fb}^{-1}$ for
high luminosity run of the LHC, for example, a sizable number of
charged Winos with $L_T\sim {\cal O}(10)$ cm will be produced.  Thus,
if we can find their tracks, it will provide an intriguing signal
which cannot be explained in the standard model.  Requiring 10 events
with $L_T\geq 50$ cm (1 m) with ${\cal L}=100\ {\rm fb}^{-1}$, for
example, the LHC can cover $M_2\leq 350$ GeV (200 GeV).  Thus, the
search for (short) tracks of heavy charged particles is strongly
suggested when other signals from superparticles cannot be discovered
at the LHC.

Even without looking for tracks of $\tilde{W}^\pm$, an excess of
mono-jet events with large missing $E_T$ may provide another signal of
the ``no singlet scenario''.  If no special effort is made to find
Wino tracks, Wino production events look like mono-jet events with
missing $E_T$ (assuming that Winos decay with $L_T\lesssim 1$ m or
so).  Detailed study of such events may give another test of the Wino
production.

The most serious standard-model background for the mono-jet event is
from the process $pp\rightarrow Zj$ followed by
$Z\rightarrow\nu\bar{\nu}$.  At the parton level, $Zj$ production is
from $q\bar{q}\rightarrow Zg$, $gq\rightarrow Zq$, and
$g\bar{q}\rightarrow Z\bar{q}$.  In order to estimate the number of
backgrounds, we calculate the cross section of these processes as
functions of $E_T$ of the jet.  The total cross section of
$pp\rightarrow Zj$ multiplied by $Br(Z\rightarrow\nu\bar{\nu})$ is
also plotted in Fig.\ \ref{fig:cs_et}.  The background cross section
is several orders of magnitude larger than the signal cross section.
Importantly, however, the cross section of the process $pp\rightarrow
Zj$ will be well understood by using the decay mode $Z\rightarrow
l^+l^-$ (with $l^\pm$ being charged lepton).  We expect that the
charged lepton pair from $Z$ can be identified and hence
$\sigma(pp\rightarrow Zj)$ will be determined with high precision.

In order to perform a statistical analysis of mono-jet + missing $E_T$
event, it is crucial to understand the total background cross section.
As well as the process $pp\rightarrow Zj$, other processes may
contribute to the background.  One of such is the process
$pp\rightarrow ZZj$ if both $Z$'s decay invisibly.  Thus, we have also
calculated the cross section of this process.  Taking account of the
branching ratio of the invisible mode of $Z$, we have found that the
background cross section from $pp\rightarrow ZZj$ is 29 -- 0.23 fb for
$E_T=100 - 500$ GeV, which is smaller than the signal cross section
when the Wino mass is smaller than $\sim 500$ GeV.  In addition, we
will obtain some information about this class of backgrounds using
leptonic decay modes of $Z$.  Thus, we consider that the process
$pp\rightarrow ZZj$ is not a serious background.  The process
$pp\rightarrow W^\pm j$ followed by leptonic decay of $W^\pm$ may also
give backgrounds if the charged lepton goes into the beam pipe.
However, the probability of charged lepton escaping into the beam pipe
may not be so large.  In addition, the number of this type of
backgrounds will be also estimated by studying mono-jet events with a
single charged lepton and missing $E_T$.  We consider that cross
sections of other processes resulting in the mono-jet + missing $E_T$
final state are also small enough to be neglected.

If the standard-model cross section of the mono-jet + missing $E_T$
event is understood precisely, an excess beyond the statistical error
may be regarded as a signal of Wino + mono-jet events.  That is, the
expected number of background events is ${\cal O}(10^7 -10^4)$ for
$E_T >100-500$ GeV of the mono-jet at a high luminosity $L = 100\ {\rm
fb}^{-1}$, for example.  Then, the number of the Wino production
events with a mono-jet becomes statistically significant when
$M_2\lesssim 400$ GeV (see Fig.\ \ref{fig:cs_et}).

It may be the case that the background cross section cannot be
determined with such a high precision because of some systematic
errors.  We will not go into the detailed study of systematic error.
Instead, we estimate how well the systematic error should be
controlled to find an anomaly; in order to cover the Wino mass of 100,
200, 300, 400, and 500 GeV, systematic error in the determination of
background cross section should be smaller than 1.9, 0.28, 0.083,
0.032, and 0.015\ \% (12, 4.2, 1.8, 0.87, and 0.45\ \%) for
$E_T^{(min)}>100$ GeV (500 GeV).

Finally, let us briefly discuss what happens if Higgsino mass (i.e.,
the $\mu$-parameter) is smaller than $\sim$ TeV.  So far, we have
considered the case where superparticles other than gauginos are as
heavy as ${\cal O}(100)$ TeV, and hence the mass difference between
charged and neutral Winos is determined mainly by the radiative
corrections from gauge-boson loops.  In this case, $\Delta M$ is given
by Eq.\ (\ref{deltaM}) and the lifetime of charged Wino is more or
less predicted.  If the $\mu$-parameter is smaller, on the contrary,
$\Delta M$ is affected by the gaugino-Higgsino mixing.  In this case,
$\Delta M$ can be enhanced or suppressed, depending on $\mu$- and
other parameters.  In particular, $\Delta M$ becomes smaller than the
pion mass and the lifetime of charged Wino becomes longer in some
parameter region.  In this case, it is much easier to find tracks of
the charged Wino since $\tilde{W}^\pm$ does not decay inside the
detector.  It should be also noticed here that such a charged Wino
with the lifetime of ${\cal O}(10^{-8})$ sec may be used as a probe of
deep interior of heavy nuclei \cite{Hamaguchi:2006vp}.

\section*{Acknowledgment}

The authors would like thank S. Asai, J. Kanzaki and T. Kobayashi for valuable
discussion and useful comments.  
M.I.  is grateful to T. Hahn and C. Schappacher for  kind explanation 
of {\it FormCalc}.
This work was supported in part by
the U.S. Department of Energy under contract number DE-AC02-76SF00515
(M.I.) and by the Grant-in-Aid for Scientific Research from the
Ministry of Education, Science, Sports, and Culture of Japan, No.\ 
15540247 (T.M.) and 14102004 (T.T.Y.).


\begin{thebibliography}{99}

\bibitem{SUSY}
  H.~P.~Nilles,
  Phys.\ Rept.\  {\bf 110} (1984) 1;
  S.~P.~Martin,
  arXiv:hep-ph/9709356.

\bibitem{Randall:1998uk}
  L.~Randall and R.~Sundrum,
  %
  Nucl.\ Phys.\ B {\bf 557}, 79 (1999).
\bibitem{Giudice:1998xp}
  G.~F.~Giudice, M.~A.~Luty, H.~Murayama and R.~Rattazzi,
  %
  JHEP {\bf 9812}, 027 (1998).
   
\bibitem{FCNC}
  F.~Gabbiani, E.~Gabrielli, A.~Masiero and L.~Silvestrini,
  Nucl.\ Phys.\ B {\bf 477}, 321 (1996).

\bibitem{Yao:2006px}
    W.~M.~Yao {\it et al.}  [Particle Data Group],
    J.\ Phys.\ G {\bf 33}, 1 (2006).

\bibitem{gravitino}
S.~Weinberg,
  Phys.\ Rev.\ Lett.\  {\bf 48} (1982) 1303;
 See, for a recent work,  
  K.~Kohri, T.~Moroi and A.~Yotsuyanagi,
  Phys.\ Rev.\ D {\bf 73} (2006) 123511,
  and references therein.

\bibitem{Fukugita:1986hr}
    M.~Fukugita and T.~Yanagida,
    Phys.\ Lett.\ B {\bf 174}, 45 (1986);
    For a recent review, 
  W.~Buchmuller, R.~D.~Peccei and T.~Yanagida,
  Ann.\ Rev.\ Nucl.\ Part.\ Sci.\  {\bf 55}, 311 (2005).

\bibitem{Leptogen+AMSB}
    M.~Ibe, R.~Kitano, H.~Murayama and T.~Yanagida,
    Phys.\ Rev.\ D {\bf 70}, 075012 (2004);
    M.~Ibe, R.~Kitano and H.~Murayama,
    Phys.\ Rev.\ D {\bf 71}, 075003 (2005).


\bibitem{Moroi:1999zb}
    T.~Moroi and L.~Randall,
    Nucl.\ Phys.\ B {\bf 570}, 455 (2000).

\bibitem{ISY}  
  M.~Ibe, Y.~Shinbara and T.~T.~Yanagida,
  arXiv:hep-ph/0608127.

\bibitem{Feng:1999fu}
  J.~L.~Feng, T.~Moroi, L.~Randall, M.~Strassler and S.~f.~Su,
  Phys.\ Rev.\ Lett.\  {\bf 83}, 1731 (1999).
  
\bibitem{Kawasaki:2006gs}
    M.~Kawasaki, F.~Takahashi and T.~T.~Yanagida,
    Phys.\ Lett.\ B {\bf 638}, 8 (2006);
  M.~Kawasaki, F.~Takahashi and T.~T.~Yanagida,
  Phys.\ Rev.\ D {\bf 74}, 043519 (2006);
  M.~Endo, M.~Kawasaki, F.~Takahashi and T.~T.~Yanagida,
  arXiv:hep-ph/0607170.

\bibitem{Wells:2004di}
  J.~D.~Wells,
  Phys.\ Rev.\ D {\bf 71}, 015013 (2005).
\bibitem{Inoue:1991rk}
  K.~Inoue, M.~Kawasaki, M.~Yamaguchi and T.~Yanagida,
  Phys.\ Rev.\ D {\bf 45}, 328 (1992).

\bibitem{Gherghetta:1999sw}
  T.~Gherghetta, G.~F.~Giudice and J.~D.~Wells,
  Nucl.\ Phys.\ B {\bf 559}, 27 (1999).


\bibitem{Heister:2002mn}
  A.~Heister {\it et al.}  [ALEPH Collaboration],
  Phys.\ Lett.\ B {\bf 533}, 223 (2002).

\bibitem{Thomas:1998wy}
  S.~D.~Thomas and J.~D.~Wells,
  Phys.\ Rev.\ Lett.\  {\bf 81}, 34 (1998).

\bibitem{ATLAS}
  ATLAS homepage,
  {\tt http://atlas.web.cern.ch}.
  
\bibitem{CMS}
  CMS homepage,
  {\tt http://cms.cern.ch}.
\bibitem{ATLASin}
ATLAS Collaboration,
CERN-LHCC-97-16;
CERN-LHCC-97-17.

\bibitem{Pumplin:2002vw}
    J.~Pumplin, D.~R.~Stump, J.~Huston, H.~L.~Lai, P.~Nadolsky 
    and W.~K.~Tung,
    JHEP {\bf 0207}, 012 (2002).
\bibitem{FA}
  T.~Hahn,
  Comput.\ Phys.\ Commun.\  {\bf 140}, 418 (2001).
  \bibitem{FC}
  T.~Hahn and M.~Perez-Victoria,
  Comput.\ Phys.\ Commun.\  {\bf 118}, 153 (1999).


\bibitem{Hamaguchi:2006vp}
  K.~Hamaguchi, T.~Hatsuda and T.~T.~Yanagida,
  arXiv:hep-ph/0607256.

\end{thebibliography}
\end{document}